\documentclass[12pt]{article}%
\usepackage{citesort}
\usepackage{amsmath}
\usepackage{graphicx}%
\usepackage{amsfonts}%
\usepackage{amssymb}
\def\ba{\begin{eqnarray}}
\def\ea{\end{eqnarray}}

\begin{document}

\title{The diquark and elastic pion-proton scattering at high energies}
\author{Adam Bzdak\thanks{Fellow of the Polish Science Foundation (FNP) scholarship
for the year 2006.}\\M.Smoluchowski Institute of Physics \\Jagellonian University, Cracow\footnote{Address: Reymonta 4, 30-059 Krakow,
Poland; e-mail: bzdak@th.if.uj.edu.pl}}
\maketitle

\begin{abstract}
Small momentum transfer elastic pion-proton cross-section at high energies is
calculated assuming the proton is composed of two constituents, a quark and a
diquark. We find that it is possible to fit very precisely the data when (i)
the pion acts as a single entity (no constituent quark structure) and (ii) the
diquark is rather large, comparable to the size of the proton.

\end{abstract}

\section{Introduction}

In this note we study the quark correlations inside the nucleon - forming a
diquark \cite{Jaffe} - in the context of elastic pion-proton scattering at low
momentum transfer.

The interest in this process is a consequence of Ref. \cite{bb2} where the
elastic proton-proton scattering, assuming the proton is composed of quark and
diquark, was discussed. We found that (i) it was possible to fit very
precisely the ISR elastic $pp$ data \cite{elastic} even up to $-t\approx3$
GeV$^{2}$ and (ii) the diquark turned out to be rather large, comparable to
the size of the proton. Moreover, we found that the quark-diquark model of the
nucleon in the wounded \cite{bbc} constituent model \cite{bb1} allows to
explain very well the RHIC data \cite{phobos} on particle production in the
central rapidity region.

Given above arguments, it is interesting to explore the model for another
process. The natural one is elastic pion-proton scattering. Following
\cite{bb2} we consider the proton to be composed of two constituents - a quark
and a diquark. As far as the pion is concerned we consider two cases. The
first one treats the pion as an object composed of two constituent quarks, the
second one treats the pion as a single object i.e. an object without
constituent quark structure.

For both cases we evaluate the inelastic pion-proton cross-section,
$\sigma(b)$, at a given impact parameter $b$. Then, from the unitarity
condition we obtain the elastic amplitude\footnote{We ignore the real part of
the amplitude.}
\begin{equation}
t_{el}(b)=1-\sqrt{1-\sigma(b)}, \label{unitarity}%
\end{equation}
and consequently the elastic amplitude in momentum transfer representation:%
\begin{equation}
T(\Delta)=\int t_{el}(b)e^{i\vec{\Delta}\cdot\vec{b}}d^{2}b.
\end{equation}

With this normalization one can evaluate the total cross section:%
\begin{equation}
\sigma_{tot}=2T(0), \label{tot}%
\end{equation}
and elastic differential cross section ($t\simeq-|\Delta|^{2}$):%
\begin{equation}
\frac{d\sigma}{dt}=\frac{1}{4\pi}|T(\Delta)|^{2}. \label{elas}%
\end{equation}

Our strategy is to adjust the parameters of the model so that it fits best the
data for elastic pion-proton cross-section. In this way the model can provide
some information on the details of proton and pion structure at small momentum transfer.

\section{Pion as a quark-quark system}

\begin{figure}[h]
\begin{center}
\includegraphics[scale=1]{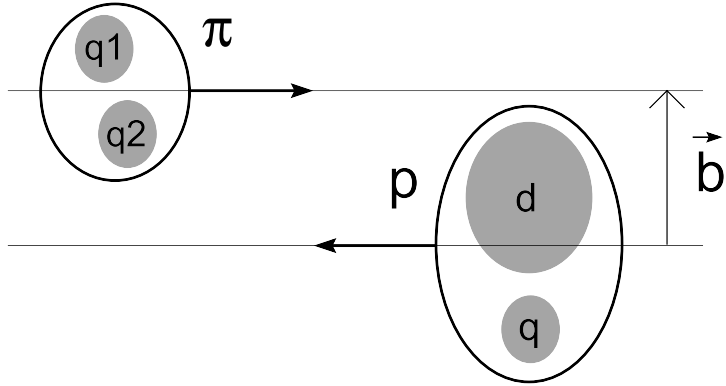}
\end{center}
\caption{Pion-proton scattering in the quark-diquark model. Pion as a
quark-quark system.}%
\label{fig1}%
\end{figure}

We follow closely the method presented in \cite{bb2} where the elastic and
inelastic proton-proton collision was studied. Consequently, the inelastic
pion-proton cross-section at a fixed impact parameter $b$, $\sigma(b)$, is
given by:%
\begin{equation}
\sigma(b)=\int d^{2}s_{q}d^{2}s_{d}d^{2}s_{q1}d^{2}s_{q2}D_{p}(s_{q}%
,s_{d})D_{\pi}(s_{q1},s_{q2})\sigma(s_{q},s_{d};s_{q1},s_{q2};b),
\label{sigma}%
\end{equation}
where $D_{p}(s_{q},s_{d})$ and $D_{\pi}(s_{q1},s_{q2})$ denote the
distribution of quark ($s_{q}$) and diquark ($s_{d}$) inside the proton and
the distribution of quarks ($s_{q1}$,$s_{q2}$) inside the pion, respectively.
$\sigma(s_{q},s_{d};s_{q1},s_{q2};b)$ is the probability of inelastic
interaction at fixed impact parameter $b$ and frozen transverse positions of
all constituents. The schematic view of this process is shown in Fig.
\ref{fig1}.

Using the Glauber \cite{Glauber} and Czyz-Maximon \cite{cm} expansions we
have:\footnote{Here and in the following we assume that all constituents act
independently.}%
\begin{align}
1-\sigma(s_{q},s_{d};s_{q1},s_{q2};b)  &  =[1-\sigma_{qq}(b+s_{q1}%
-s_{q})][1-\sigma_{qq}(b+s_{q2}-s_{q})]\times\nonumber\\
\times &  [1-\sigma_{qd}(b+s_{q1}-s_{d})][1-\sigma_{qd}(b+s_{q2}-s_{d})],
\end{align}
where $\sigma_{ab}(s)$ ($ab$ denotes $qq$ or $qd$) are inelastic differential
cross-sections of the constituents. 

Following \cite{bb2} we parametrize $\sigma_{ab}(s)$ using simple Gaussian
forms:%
\begin{equation}
\sigma_{ab}(s)=A_{ab}e^{-s^{2}/R_{ab}^{2}}. \label{pab}%
\end{equation}
We constrain the radii $R_{ab}$ by the condition: $R_{ab}^{2}=R_{a}^{2}%
+R_{b}^{2}$ where $R_{a}$ denotes the quark or diquark's radius.\footnote{We
assume the quark radius to be the same in the proton and pion. We have
checked, however, that allowing the $R_{q}$'s to vary independently we are led
to the same conclusions.}

From (\ref{pab}) we obtain the total inelastic cross sections: $\sigma
_{ab}=\pi A_{ab}R_{ab}^{2}$. Following \cite{bb2} we assume that the ratios of
cross-sections satisfy the condition: $\sigma_{qd}/\sigma_{qq}=2$, what allows
to evaluate $A_{qd}$ in term of $A_{qq}$.

For the distribution of the constituents inside the proton we take a Gaussian
with radius $R$:%
\begin{equation}
D_{p}(s_{q},s_{d})=\frac{1+\lambda^{2}}{\pi R^{2}}e^{-(s_{q}^{2}+s_{d}%
^{2})/R^{2}}\delta^{2}(s_{d}+\lambda s_{q}), \label{Dp}%
\end{equation}
where the parameter $\lambda$ has the physical meaning of the ratio of the
quark and diquark masses $\lambda=m_{q}/m_{d}$ (the delta function guarantees
that the center-of-mass of the system moves along the straight line). One
expects $1/2\leq\lambda\leq1$.

For the distribution of quarks inside the pion we take a Gaussian with radius
$d$:
\begin{equation}
D_{\pi}(s_{q1},s_{q2})=\frac{1}{\pi d^{2}}e^{-(s_{q1}^{2}+s_{q2}^{2})/2d^{2}%
}\delta^{2}(s_{q1}+s_{q2}). \label{Dpi}%
\end{equation}
It allows to define the effective pion radius $R_{\pi}$:%
\begin{equation}
R_{\pi}^{2}=d^{2}+R_{q}^{2}.
\end{equation}

Now the calculation of $\sigma(b)$, given by (\ref{sigma}), reduces to
straightforward Gaussian integrations. The relevant formula is given in the
Appendix. Introducing this result into the general formulae given in Section
$1$ one can evaluate the total and elastic differential pion-proton cross-sections.

\begin{figure}[h]
\begin{center}
\includegraphics[scale=1.4]{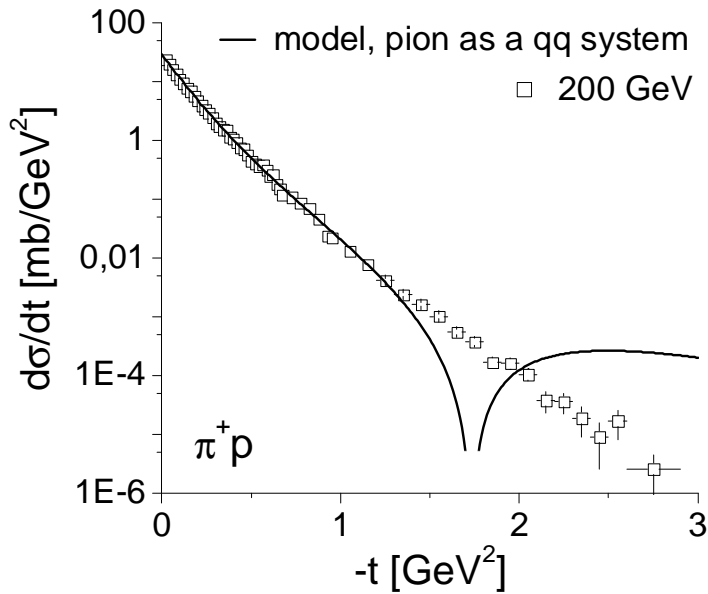}
\end{center}
\caption{The model compared to data \cite{pion} on differential cross section
for elastic $\pi^{+}p$ scattering at $p_{lab}=200$ GeV (not all points
plotted, for clarity). Pion as a quark-quark system.}%
\label{pip1}%
\end{figure}

Our strategy is to adjust the parameters of the model so that it fits the data
best. We have analyzed the data for elastic $\pi^{+}p$ scattering at two
incident momenta $p_{lab}=$ $100$ GeV and $200$ GeV \cite{pion}. An example of
our calculation is shown in Fig. \ref{pip1}, where the differential cross
section $d\sigma/dt$ at $p_{lab}=200$ GeV, evaluated from the model, is
compared with data \cite{pion}.

From Fig. \ref{pip1} one sees that it is possible to fit very precisely the
data up to $-t\approx1$ GeV$^{2}$. However, the model, with the pion as a
two-quark system, predicts the diffractive minimum which is not seen in the
data. In the next section we show that assuming the pion to be a single entity
(no constituent quark structure) we are able to remove this problem.

The relevant values of the parameters are given in Table \ref{Tab_a}%
.\footnote{The model is almost insensitive to the value of $\lambda$ (provided
that $1/2\leq$ $\lambda\leq1$).} \begin{table}[h]
\begin{center}%
\begin{tabular}
[c]{|c|c|c|c|c|c|}\hline\hline
$p_{lab}$ [GeV] & $R_{q}$ [fm] & $R_{d}$ [fm] & $R$ [fm] & $R_{\pi}$ [fm] &
$A_{qq}$\\\hline
$200$ & 0.26 & 0.82 & 0.28 & 0.44 & 1\\\hline
\end{tabular}
\end{center}
\caption{The parameters of the model at the incident momentum $p_{lab}=200$
GeV. Pion as a quark-quark system.}%
\label{Tab_a}%
\end{table}

It is intriguing to notice that the values of the most interesting parameters,
$R_{q}$ and $R_{d}$, are not far from those obtained in \cite{bb2} were
elastic $pp$ scattering was studied in similar approach. Again we observe that
the diquark is rather large.

\section{Pion as a single entity}

\begin{figure}[h]
\begin{center}
\includegraphics[scale=1]{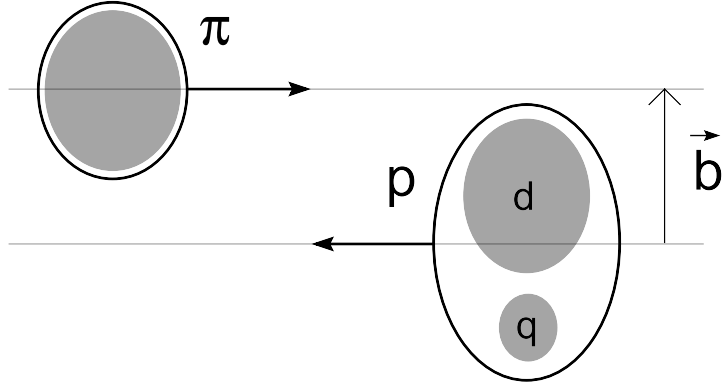}
\end{center}
\caption{Pion-proton scattering with the pion as a single entity.}%
\label{fig2}%
\end{figure}

In the present section we assume that the pion interacts as a single entity
i.e. the pion has no constituent quark structure. The schematic view of the
pion-proton scattering in this approach is shown in Fig. \ref{fig2}.

The inelastic pion-proton cross-section $\sigma(b)$ at a fixed impact
parameter $b$ reads:%
\begin{equation}
\sigma(b)=\int d^{2}s_{q}d^{2}s_{d}D_{p}(s_{q},s_{d})\sigma(s_{q},s_{d};b),
\end{equation}
with $D_{p}(s_{q},s_{d})$ given by (\ref{Dp}) and $\sigma(s_{q},s_{d};b)$
expressed by:
\begin{equation}
1-\sigma(s_{q},s_{d};b)=[1-\sigma_{q\pi}(b-s_{q})][1-\sigma_{d\pi}(b-s_{d})].
\end{equation}

In analogy to the previous approach the inelastic differential quark-pion,
$\sigma_{q\pi}(s)$, and diquark-pion, $\sigma_{d\pi}(s)$, cross sections are
parametrized using simple Gaussian:%
\begin{equation}
\sigma_{a\pi}(s)=A_{a\pi}e^{-s^{2}/R_{a\pi}^{2}}. \label{p_a_pi}%
\end{equation}
In this case we constrain the radii $R_{a\pi}$ by the condition: $R_{a\pi}%
^{2}=R_{a}^{2}+R_{\pi}^{2}$ where $R_{a}$ denotes the quark or diquark's
radius and $R_{\pi}$ denotes the pion's radius.

This gives:%
\begin{align}
\sigma(b)  &  =\frac{A_{q\pi}x}{x+r}e^{-b^{2}/(x+r)}+\frac{A_{d\pi}%
y}{y+\lambda^{2}r}e^{-b^{2}/(y+\lambda^{2}r)}\\
&  -\frac{A_{q\pi}A_{d\pi}xy}{xy+yr+\lambda^{2}xr}e^{-b^{2}\tfrac
{x+y+(1+\lambda)^{2}r}{xy+yr+\lambda^{2}xr}},\nonumber
\end{align}
where $x=R_{q}^{2}+R_{\pi}^{2},$ $y=R_{d}^{2}+R_{\pi}^{2}$ and $r=R^{2}%
/(1+\lambda^{2})$. Introducing this result into the general formulae given in
Section $1$ one can evaluate the total and elastic differential pion-proton cross-sections.

From (\ref{p_a_pi}) we deduce the total inelastic cross sections:
$\sigma_{a\pi}=\pi A_{a\pi}R_{a\pi}^{2}$. As before we demand that the ratios
of cross-sections satisfy the condition: $\sigma_{d\pi}/\sigma_{q\pi}=2$, what
allows to evaluate $A_{d\pi}$ in term of $A_{q\pi}$.

\begin{figure}[h]
\begin{center}
\includegraphics[scale=1.4]{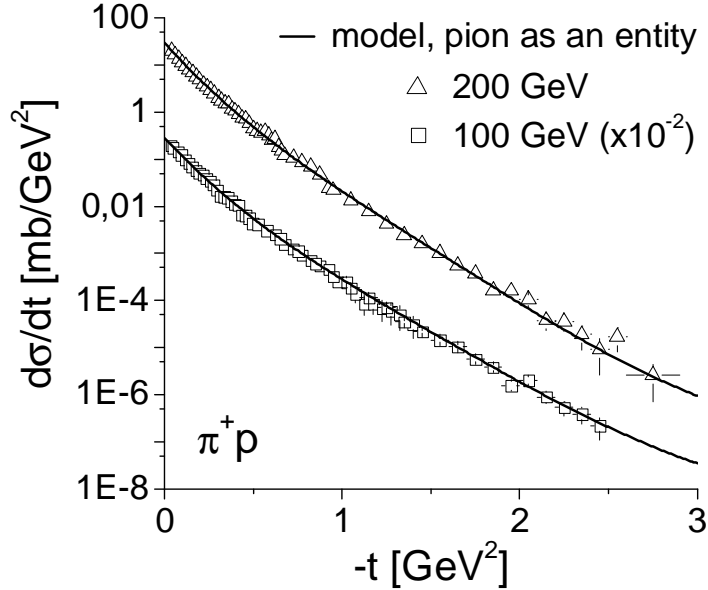}
\end{center}
\caption{Pion acts as a single entity. The model compared to data \cite{pion}
(not all points plotted, for clarity) on differential cross section for
elastic $\pi^{+}p$ scattering at $p_{lab}=100$ GeV (rescaled by a factor
$10^{-2}$) and $200$ GeV.}%
\label{pip2}%
\end{figure}

It turns out that the model in this form works very well indeed i.e. it is
possible to fit very precisely the data even up to $-t\approx3$ GeV$^{2}$. We
have analyzed the data at two incident momenta of $100$ and $200$ GeV
\cite{pion}. The results of our calculations are shown in Fig. \ref{pip2}%
.

The relevant values of the parameters are given in Table \ref{Tab_b}. Again
the most interesting observation is the large size of the diquark, comparable
to the size of the proton. \begin{table}[h]
\begin{center}%
\begin{tabular}
[c]{|c|c|c|c|c|c|}\hline\hline
$p_{lab}$ [GeV] & $R_{q}$ [fm] & $R_{d}$ [fm] & $R$ [fm] & $R_{\pi}$ [fm] &
$A_{q\pi}$\\\hline
$100$ & 0.25 & 0.79 & 0.28 & 0.50 & 0.80\\\hline
$200$ & 0.25 & 0.79 & 0.28 & 0.52 & 0.75\\\hline
\end{tabular}
\end{center}
\caption{The parameters of the model at two incident momenta $p_{lab}=100$ GeV
and $200$ GeV. Pion acts as a single entity. }%
\label{Tab_b}%
\end{table}

\section{ Discussion and conclusions}

In conclusion, it was shown that the constituent quark-diquark structure of
the proton can account very well for the data on elastic $\pi^{+}p$
scattering. The detailed confrontation with data allows to determine the
parameters characterizing the proton and pion structure. We confirm the large
size of the diquark, while the pion seems to interact as a single entity i.e.
without constituent quark structure.

Several comments are in order.

(a) We compared the model only to elastic $\pi^{+}p$ scattering data, however,
there is no statistically significant difference between $\pi^{+}p$ and
$\pi^{-}p$ data \cite{pion} at any $t$ value (at least up to $-t\approx3$
GeV$^{2}$).

(b) The pion seems to interact as a single entity. It suggests that during
pion-nucleus collision the pion produces the same number of particles no
matter how many inelastic collisions it undergoes.\footnote{We thank Robert
Peschanski for this observation.}

\bigskip

\textbf{Acknowledgements}

We thank Andrzej Bialas for suggesting this investigation and illuminating
discussions. Discussions with Stephane Munier, Robert Peschanski, Michal
Pra{\-}sza{\-}lo{\-}wicz and Samuel Wallon are also highly appreciated.

\section{Appendix}

The problem is to calculate the following integral:%
\begin{gather}
\frac{r_{1}r_{2}}{\pi^{2}}\int d^{2}sd^{2}s^{\prime}e^{-r_{1}s^{2}}%
e^{-r_{2}s^{\prime2}}e^{-y_{1}(b+\lambda s-s^{\prime})^{2}}e^{-y_{2}(b+\lambda
s+s^{\prime})^{2}}\times\\
\times e^{-x_{1}(b-s-s^{\prime})^{2}}e^{-x_{2}(b-s+s^{\prime})^{2}%
}=\frac{r_{1}r_{2}}{\Omega}e^{-b^{2}\Gamma/\Omega},\nonumber
\end{gather}
where:%
\begin{align}
\Gamma &  =(1+\lambda)^{2}\left[  4x_{1}x_{2}(y_{1}+y_{2})+4y_{1}y_{2}%
(x_{1}+x_{2})+r_{2}(x_{1}+x_{2})(y_{1}+y_{2})\right] \nonumber\\
&  +4r_{1}(x_{1}+y_{1})(x_{2}+y_{2})+r_{1}r_{2}(x_{1}+x_{2}+y_{1}+y_{2}),
\end{align}%
\begin{align}
\Omega &  =(1-\lambda)^{2}(x_{1}y_{2}+x_{2}y_{1})+(1+\lambda)^{2}(x_{1}%
y_{1}+x_{2}y_{2})+\lambda^{2}[r_{2}(y_{1}+y_{2})+4y_{1}y_{2}]\nonumber\\
&  +r_{1}(x_{1}+x_{2}+y_{1}+y_{2}+r_{2})+r_{2}(x_{1}+x_{2})+4x_{1}x_{2}.
\end{align}

Other needed integrals can be obtained by putting some of the $x_{1},$
$x_{2},$ $y_{1}$ or $y_{2}=0$.

\end{document}